\documentclass[aps,pra,onecolumn,groupedaddress,showpacs]{revtex4}

\usepackage{hyperref}

\usepackage{amssymb,amsmath}

\draft

\usepackage{graphicx}
\newcommand{\Om}{\Omega}

\newcommand{\om}{\omega}
\newcommand{\al}{\alpha}

\newcommand{\ve}{\varepsilon}
\newcommand{\pa}{\partial}

\begin{document}

\title{Vector Pulsing Solitons in Semiconductor Quantum Dots}

\author{G. T. Adamashvili, M. D. Peikrishvili, R. R. Koplatadze, K.L.Schengelia}
\affiliation{Technical University of Georgia, Kostava str. 77, 0179, Tbilisi,
Georgia \\ email: $guram_{-}adamashvili@ymail.com$}

\begin{abstract}
A theory of an optical vector pulsing soliton of self-induced transparency in an ensemble of semiconductor quantum dots is investigated. It is shown that a distribution of the excitonic ground-state transition dipole moments of the quantum dots  and phase modulation changes significantly the pulse parameters. The profile of a circularly polarized optical two-component vector pulsing soliton with the difference and sum of the frequencies in the region of the carrier frequency is presented. It is shown that the vector pulsing soliton in the special case  can be reduced to the scalar breather solution and these nonlinear waves have different shapes. Explicit analytical expressions for the optical vector pulsing soliton are obtained with realistic parameters which can be reached in current experiments.
\end{abstract}


\maketitle

\centerline{I. Introduction}

Semiconductor quantum dots (SQDs) as model systems for investigation of nonlinear light-matter interaction have attracted much interest.
SQDs, also referred to as zero-dimensional systems, that allow confinement of  charge carriers in all three spatial directions, which results in atomic-like discrete energy spectra and  strongly enhanced carrier lifetimes \cite{Bimberg::1999}. Due to the large transition dipole moments of quantum dots the interaction between SQDs and optical pulse is strongly enhanced in comparison with atomic systems, making them especially attractive for nonlinear nano-optics and different applications. In addition, to very long relaxation times in quantum dots are of order of several tens of picoseconds \cite{Borri::02, Panzarini:PhysRevB:02}, optical wave propagation experiments can be performed with pulses of a few picoseconds. Study of optical effects in ensembles of SQDs is usually spoiled by the inhomogeneous line broadening due to dot-size fluctuations, with a full width at half maximum of typically more than several tens of meV. Real quantum dots often have a base length in the range 50-400~{\AA}. Beside  the frequency, the quantum dot-size fluctuations influence also the transition dipole moments of SQDs $\vec{\mu}$. Borri {\it  et al.} \cite{ Borri::02} have reported measurements of optical Rabi oscillations in the excitonic ground-state transition  of an inhomogeneously broadened InGaAs quantum dot ensemble. They found that a distribution with a 20 percent standard deviation of transition dipole moments results in a strong damping of the oscillations versus pulse area.
In the experiments reported in Ref.\cite{Borri::02} it was also  found that  the period of the Rabi oscillations is changed. These results show quantitatively how uniformity in dot size is important for any application based on a coherent light-quantum dot ensemble interaction.

A nonlinear coherent interaction of an optical pulse with quantum dots, is governed by the Maxwell-Bloch equations. The large numerical values of the transition dipole moments and very wide inhomogeneous broadening of spectral line do not change the Maxwell-Bloch equations for SQDs in comparison with atomic systems. But considering the distribution of the transition dipole moments of SQDs, the polarization and Maxwell equation of the SQDs  will be differ in comparison with atomic systems  because the polarization depends from the variable $\vec{\mu}$ \cite{AdamshviliMaradudin:Arxiv:2010}.

The formation of nonlinear waves is one of the most interesting manifestations of nonlinearity in ensembles of quantum dots. The resonant optical nonlinear wave can be formed with the help of the resonance (McCall-Hahn) mechanism of the formation of nonlinear waves, i.e. from a nonlinear coherent interaction of an optical pulse with resonant ensembles of SQDs, when the conditions of the self-induced transparency (SIT): $\omega T>>1$ and $T<<T_{1,2}$ are fulfilled, where $T$ and $\omega $ are the width and frequency of the pulse, and $T_1$ and $T_2$ the relaxation and dephasing time of SQDs, respectively \cite{Panzarini:PhysRevB:02, Allen::75}.
If the area of the pulse envelope as a measure of the light-matter interaction strength $\Theta >\pi$, a soliton is formed, and if $\Theta<<1$, a breather (pulsing soliton) is generated. The importance of breathers is that they have many soliton-like properties, but unlike solitons can be excited with relatively small intensity of input pulses which makes the experimental realization easier. In addition, in some situations when soliton are unstable, the instability almost always leads to the formation of a breather \cite{Chen:Phys. Rev. E :04}.
Thus, in some cases, breathers seem to be more stable and numerous than solitons.

Experimentally, self-induced transmission on a free exciton resonance in CdSe \cite{Giessen:PhysRevLett:98} and SIT in InGaAs quantum dot waveguides have been reported \cite{Schneider:ApplPhysLett:03}. On the theoretical side, the effect of SIT in SQDs have been investigated \cite{Panzarini:PhysRevB:02, Adamashvili:OptLett:06, Adamshvili:PhysRevA:07, Adamashvili:PhysLettA:07, Adamshvili:Eur.Phys.J.D.:08, Adamashvili:Opt.and Sprct:09, Adamshvili:Result:11}. In these works breathers are described by a single nonlinear Schr\"odinger (NLS) equation for a scalar field. Such scalar nonlinear waves form when a single pulse propagates inside a medium containing SQDs  in such a way that it maintains its state. When these conditions are not satisfied, one have to consider interaction of two field components at different frequencies or polarizations and solve simultaneously a set of coupled NLS equations. A profile-preserving solution of these equations is a vector  pulse  because of its two-component configuration.

The main goal of this work is as follows. The investigation of the processes of the formation of a circularly polarized optical vector pulsing soliton with two different frequency components in the region of the carrier frequency. The determination of the explicit analytic expressions for the parameters  of the two-component vector pulsing soliton with the difference and sum  of the frequencies for the strength of the electric field of the wave. The analytical solution of the Maxwell-Bloch equations for the ensemble of SQDs and explain how these equations are modified in the case of  distribution of the dipole moments and phase modulation.
To prove that influence of phase modulation on SIT in SQDs do the conditions of existence  of vector pulsing soliton less hard in comparison with conditions of  existence of vector pulses SIT investigated earlier \cite {Adamshvili:Result:11}.

\vskip+0.5cm

\centerline{II. Basic equations}

We consider the propagation of a circularly polarized optical coherent plane wave pulse in an ensemble of SQDs along the positive $z$ axis, with a  pulse  width $T<<T_{1,2}$, frequency $\omega >>T^{-1}$, wave vector $\vec{k}$, and the  strength of the optical electrical field
\begin{equation}\label{eq1}
\vec{E}(z,t)=\sum_{l=\pm 1}\vec{e}_{l}\hat{E}_{l} Z_{l}
\end{equation}
where
\begin{equation}\label{eq2}\nonumber\\
\hat{E}_{l}=\hat{E} e^{-il \phi}
,\;\;\;\;\;\;\vec{e}_{l}=\frac{1}{\sqrt{2}}(\vec{x}+l i\vec{y})
\end{equation}
is the complex polarization vectors, $\vec{x}$ and $\vec{y}$ are unit vectors in the directions of the $x$ and $y$ axes,
$Z_{l}=e^{il(kz-\om t)}$, $\hat{E}(z,t)$ is the slowly changing real envelope of the electric field, $\phi(z,t)$ is the phase function.  To guarantee the reality of the quantity $\vec{E}(z,t)$ we set $\hat{E}_{l}=\hat{E}^{*}_{-l}.$

We use the method of slowly varying envelopes, and assume that envelope $\hat{E}_l$   satisfied
the conditions
\begin{equation}\label{eq3}\nonumber\\
|\frac{\partial \hat{E}_l}{\partial t}|<<\omega
|\hat{E}_l|,\;\;\;|\frac{\partial \hat{E}_l}{\partial z }|<<k|\hat{E}_l|.
\end{equation}

For the description of interaction of quantum dot in a circularly polarized optical pulse
we use the optical selection rules for quantum dot and consider two-level system consisting of the ground state $|1>$  and one exciton state $|2>$ (0-X transitions)
with energies $\mathcal{E}_{1}=0$ and $\mathcal{E}_{2}=\hbar \omega_{0}$, respectively \cite{Bimberg::1999, Borri::02, Panzarini:PhysRevB:02}. Here, $\hbar$ is Planck's constant, $ \omega_{0}$ is the frequency of excitation of the two-level quantum dot.

The Hamiltonian and wave function of this system are:
\begin{equation}\label{eq4}
H=H_{0}+\hat V,
\end{equation}
$$
|\Psi >=\sum_{n=1,2} c_{n}(t) e^{-\frac{i}{\hbar} \mathcal{E}_{n} t }|n>,
$$
where $H_{0}=\hbar\omega_{0}|2><2|$ is the Hamiltonian of the two-level SQD,
$$
\hat V=-\hat{\mu} \vec{E}=-\mu (\hat{\sigma}_{-} E_{+1}+\hat{\sigma}_{+}E_{-1} )
$$
is the light-SQD interaction Hamiltonian,
$
 \hat{\mu}=\vec{\mu}_{12}  \hat{\sigma}_{+}+ \vec{\mu}^{*}_{12}\hat{\sigma}_{-}
$
is the SQD's dipole moment operator,
$$
\vec{\mu}_{12}=\frac{\mu}{\sqrt{2}}(\vec{x} - i \vec{y})
$$
is the electric dipole matrix element for the corresponding transition,
$
\mu=|\vec{\mu}_{12}|,\;\;\vec{\mu}_{21}=\vec{\mu}^{*}_{12},
$
$$
{E}_{\pm 1} =\hat{E}_{\mp 1} Z_{\mp 1},
$$
$$
\hat{{\sigma}}_{\pm}=\frac{1}{2}(\hat{\sigma}_{x}\pm i\hat{\sigma}_{y}),
$$
and the $\{\hat{{\sigma}}_{i}\}$ are the Pauli matrices, which satisfy $[\hat{{\sigma}}_{x},\hat{{\sigma}}_{y}]=2i\hat{{\sigma}}_{z}$, and commutation relations resulting from cyclic permutations of the subscripts (where $i=x,y,z$).

The probability amplitudes $c_{1}$ and $c_{2}$ are determined by the Schr\"odinger equations \cite{Landau:Quantum Mechanics:80}:
\begin{equation}\label{schr}
i\hbar \frac{\partial c_{1}(t)}{\partial t} = -\mu \;{E}_{+1} c_{2}(t)e^{-i\om_{0}t},
$$$$
i\hbar \frac{\partial c_{2}(t)}{\partial t} =-\mu\; {E}_{-1} c_{1}(t)e^{i\om_{0}t}.
\end{equation}

From the Hamiltonian \eqref{eq4} we obtain the Bloch equations \cite{Panzarini:PhysRevB:02,AdamshviliMaradudin:Arxiv:2010, Allen::75}:

\begin{equation}\label{equ5}
\dot{\hat{\sigma}}_{+}=i \omega_{0} \hat{\sigma}_{+}+i \frac{\mu}{\hbar} \hat{\sigma}_{z}E_{+1},
$$
$$
 \dot{\hat{\sigma}}_{-}=-i\omega_{0}\hat{\sigma}_{-} -i \frac{\mu}{\hbar} \hat{\sigma}_{z}E_{-1},
$$
$$
\dot{\hat{\sigma}}_{z}=i \frac{2 \mu}{\hbar} (\hat{\sigma}_{+}E_{-1}-\hat{\sigma}_{-} E_{+1}).
\end{equation}

The average values of the Pauli operators $\hat{\sigma}_{i}$ for the state $|\Psi >=c_{1}\;|1> + \; c_{2}\;|2>$, are $s_{i} = <\hat{\sigma}_{i}>\;\;$ $ = <\Psi|\hat{\sigma}_{i}|\Psi> $, and have the form \cite{Landau:Quantum Mechanics:80}:
\begin{equation}\label{equ6}
s_{x}=c^{*}_{1}(t)c_{2}(t)e^{-i\om_{0}t}+c_{1}(t)
c^{*}_{2}(t)e^{i\om_{0}t},
$$$$
s_{y}= ic^{*}_{1}(t)c_{2}(t)e^{-i\om_{0}t}
-ic_{1}(t)c^{*}_{2}(t)e^{i\om_{0}t},
$$$$
s_{z}=c^{*}_{2}(t) c_{2}(t)-c^{*}_{1}(t)c_{1}(t).
\end{equation}

Unlike the Bloch equations for atomic systems, in the Eqs. \eqref{equ5} for SQDs the quantity $\mu$ is
not constant and for different quantum dots has different value. Consequently the probability amplitudes $c_{1},\;c_{2}$ [Eq.\eqref{schr}] and the
quantities $s_{x},\;s_{y}$ and $s_{z}$ [Eq.\eqref{equ6}] are also functions of the variable $\mu$.

The system of equations \eqref{equ5} are exact only in the
limit of infinite relaxation times. To take into account
that we consider a coherent interaction of the optical pulse with
quantum dots, i.e. $T\ll T_{1,2}$, the influence of relaxation on the
nonlinear wave processes are neglected in
the present work.

In addition to the Eqs. \eqref{equ5}, we need a description of the pulse propagation in the medium. The wave equation for the electric field $\vec{E}(z,t)$
of the optical pulse in the medium is given by
\begin{equation}\label{eq7}
\frac{\pa^{2} \vec{ E}}{\pa {z}^2} -\frac{\eta^{2}}{c^2} \frac{\pa^{2} \vec
{E}}{\pa t^2}=\frac{4\pi}{c^2}\frac{\pa^{2}\vec P }{\pa t^2},
\end{equation}
where $c$ is the speed of  light  in vacuum, $\eta$ is the refractive index of the medium, and
$\vec{P}$ is the polarization of the ensemble of  SQDs.

For the determination of the polarization of the ensemble of  SQDs we have to take into account that the dipole moment of the SQD depends on the size of the quantum dot. The polarization of the ensemble of  SQDs is equal to
\begin{equation}\label{eqy8}
\vec{P}(z,t)=\frac{n_{0}}{2}\sum_{l=\pm 1}\;\vec{e}_{l} Z_{l} \langle \mu\;p_{l}\;\rangle,
\end{equation}
where
\begin{equation}\label{we4}
 \langle \mu\;p_{l}\;\rangle = \int \int g(\Delta ) h(\mu_{0}-\mu )\mu \; p_{l}(\Delta,\mu ,z,t)\;d \mu \; d \Delta,
\end{equation}
$n_{0}$ is the uniform dot density, $g(\Delta)$ is the inhomogeneous broadening lineshape function, $\Delta=\omega_{0}-\omega.$ $p_{l}$ is the slowly changing complex amplitude of the polarization, $h(\mu_{0}-\mu)$ is the distribution function of transition dipole moments of SQDs. For this function  the normalization condition has the form
$$
\int_{-\infty}^{\infty}{h}(\mu-{\mu}_{0})d \mu =1,
$$
where
$$
\vec{\mu}_{0}=\frac{\sum_{i=1}^{n_{0}} \vec{\mu}_{i}}{n_{0}}
$$
is the main dot dipole matrix element (for details, see Ref.\cite{AdamshviliMaradudin:Arxiv:2010}).
Eq. \eqref{eqy8} generalizes the polarization of the ensemble of SQDs to the case of a distribution of transition dipole moments.  In the particular case,  if we ignore a fluctuation of the dipole moments in an ensemble of SQDs,  the quantity $\mu$ is constant, all  equations will be simplified if  the function of the distributions of the dipole moments of an SQDs $h(\mu-\mu_{0})$  will be replaced by the Dirac  $\delta$-function $\delta(\mu-\mu_{0})$ and than  we obtain all earlier known results \cite{Panzarini:PhysRevB:02,Allen::75, Adamashvili:OptLett:06, Adamshvili:PhysRevA:07, Adamashvili:PhysLettA:07, Adamshvili:Eur.Phys.J.D.:08, Adamashvili:Opt.and Sprct:09, Adamshvili:Result:11}.

By substituting Eqs. \eqref{eq1} and \eqref{eqy8} into the wave equation \eqref{eq7} we obtain  wave equations for
the slowly changing complex amplitudes $\hat{E}_{l}$  and $p_{l}$ in the following forms:
\begin{equation}\label{rty2}
\sum_{l=\pm 1}\;\vec{e}_{l} Z_{l}(-2il \om  \frac{\pa \hat{E}_{l}}{\pa t}-2ik l v^{2} \frac{\pa \hat{E}_{l}}{\pa z}
+ \frac{\pa^{2} \hat{E}_{l}}{\pa t^{2}}- v^{2} \frac{\pa^{2} \hat{E}_{l}}{\pa z^{2}}$$$$
-  \frac{ 2 \pi {\om}^{2}n_{0} }{\eta^{2}}\;\langle \mu\;p_{l}\;\rangle )=0,
\end{equation}
where  $v=c/\eta$ is the speed of light in medium.
Eqs. \eqref{we4} and \eqref{rty2} are the general equations for the slowly varying complex amplitudes $p_{l}$ and $\hat{E}_{l}$   by means of which we can consider a quite wide class of coherent optical phenomena
(for instance: Rabi oscillations,  photon echo, self-induced transparency, and others)
in the presence of phase modulation and a distribution of transition dipole moments in an ensemble of SQDs.

To further analyze of these equations we make use of the multiple scale perturbative reduction method \cite{Taniuti::1973}, in the limit that $\Theta_{l} (z,t)={\kappa}_{0}\;\vartheta_{l}(z,t)$ is ${\cal O}(\epsilon)$, with its scale-length being of order ${\cal O}(\epsilon^{-1})$, where ${\kappa}_{0}=\frac{2\mu }{\hbar}$, $\vartheta_{l}(z,t)=\int_{-\infty}^{t}\hat{E}_{l}(z,t')dt'$. This is the typical scaling for the coupled NLS equations and  would then also be the scaling for two-component pulsing soliton. In this case $\vartheta_{l}(z,t)$ can be represented as:
\begin{equation}\label{rty3}
\vartheta_{l} (z,t)=  \sum_{\al=1}^{\infty} \ve^\al {{\vartheta}_{l}}^{(\al)}(z,t)=
\end{equation}
\begin{equation}\label{rty32}
\sum_{\alpha=1}^{\infty}\sum_{n=-\infty}^{+\infty}\varepsilon^\alpha
Y_{l,n} \varphi_{l,n}^ {(\alpha)}(\zeta_{l,n},\tau),
\end{equation}
where
$$
Y_{l,n}=e^{in(Q_{l,n}z-\Omega_{l,n}
t)},
$$$$
\zeta_{l,n}=\varepsilon Q_{l,n}(z-{v_g}_{(l,n)}
t),
\;\;
\tau=\varepsilon^2 t,\;\;
{v_g}_{(l,n)}=\frac{d\Omega_{l,n}}{dQ_{l,n}},
$$
$\varepsilon$ is a small parameter. Such a representation allows
us to separate from $\vartheta_{l}$ the still more slowly changing
quantities $ \varphi_{l,n}^{(\alpha )}$. Consequently, it is assumed that
the quantities $\Omega_{l,n}$, $Q_{l,n}$, and $\varphi_{l,n}^{(\alpha)}$ satisfy the
inequalities for any $l$ and $n$:
\begin{equation}\label{rtyp}
\omega\gg \Omega,\;\;k\gg Q,\;\;\;
\end{equation}
$$
\left|\frac{\partial
\varphi_{l,n}^{(\alpha )}}{
\partial t}\right|\ll \Omega \left|\varphi_{l,n}^{(\alpha )}\right|,\;\;\left|\frac{\partial
\varphi_{l,n}^{(\alpha )}}{\partial z }\right|\ll Q\left|\varphi_{l,n}^{(\alpha )}\right|.
$$
We have to note that the quantities  $Q$  and $\Omega$ depends from
$l$ and $n$, but for simplicity, we omit these indexes in equations where this will not bring about mess.

Substituting Eqs.\eqref{we4} and \eqref{rty3} into Eq.\eqref{rty2}, we obtain
\begin{equation}\label{reu17}
\sum_{\alpha=1}\sum_{l=\pm1} \ve^{\alpha} Z_l [-2il\omega \frac{\partial^{2}\vartheta^{(\alpha)}_{l} }{\partial
t^{2}}- 2ilkv^{2} \frac{\partial^{2} \vartheta^{(\alpha)}_{l} }{{\partial
z}{\partial t}}+\frac{\partial^{3}\vartheta^{(\alpha)}_{l} }{\partial t^{3}}$$$$
-v^{2} \frac{\partial^{3}\vartheta^{(\alpha)}_{l} }{{\partial z^{2}}{\partial
t}})]= i \sum_{l=\pm 1}l Z_l [\alpha^{2}_{0}(\ve^{1}
 {{\vartheta}_{l}}^{(1)}
 +\ve^{2} {{\vartheta}_{l}}^{(2)}
 +\ve^{3} {{\vartheta}_{l}}^{(3)})$$$$
 -\ve^{3}
\frac{\beta^{2}_{0}}{2} \int\frac{\pa {{\vartheta}_{l}}^{(1)}}{\pa
t}{{\vartheta}_{-l}}^{(1)} {{\vartheta}_{l}}^{(1)}dt']
 +O(\ve^4),
\end{equation}
where
\begin{equation}\label{teu17}
\alpha^{2}_{0}= \frac{4\pi{\omega}^{2}n_0}{\eta^{2} \hbar}  \int \frac{g(\Delta)d\Delta}{1+\Delta^{2}T^{2}}  \int h(\mu_{0}-\mu)\mu^{2} d\mu, $$$$
\beta^{2}_{0}=\frac{16 \pi{\omega}^{2}n_0 }{\eta^{2} \hbar^{3}} \int \frac{g(\Delta)d\Delta}{1+\Delta^{2}T^{2}}  \int h(\mu_{0}-\mu)\mu^{4} d\mu.
\end{equation}

\vskip+0.5cm

\centerline{III. Vector pulsing soliton in SQDs}

On substituting Eq.\eqref{rty32}  into Eq.\eqref{reu17},  we obtain the nonlinear wave equation
\begin{equation}\label{teq17}
\sum_{l=\pm1}\sum_{\alpha=1}\sum_{n=-\infty}^{+\infty}
  \ve^\alpha  Z_{l} Y_{l,n}\{\tilde{W}_{l,n} +\ve J_{l,n}\frac{\partial }{\partial \zeta}
 +\ve^2 h_{l,n} \frac{\partial }{\partial \tau}+$$$$
{\ve}^{2}i H_{l,n} \frac{\partial^{2} }{\partial
\zeta^{2}}\}\varphi_{l,n}^ {(\alpha)}$$$$=-\ve^{3}i \frac{\beta^{2}_{0}}{2}\sum_{l=\pm 1}l Z_l
 \int\frac{\pa {{\vartheta}_{l}}^{(1)}}{\pa
t}{{\vartheta}_{-l}}^{(1)} {{\vartheta}_{l}}^{(1)}dt'
 +O(\ve^4)
\end{equation}
where
\begin{equation}\label{e15}
{W}_{l,n}=in \Omega (A_{l}n {\Omega} - B_{l} n Q  + n^{2}
{\Omega}^{2}-v^{2} n^{2} Q^{2}  -\frac{l}{n }\frac{\alpha^{2}_{0}}{ \Omega}
),
$$
$$
 J_{l,n}= n Q[2A_{l} \Omega
v_g -B_{l}(Q v_g +\Omega) + 3n {\Omega}^{2}  v_g $$$$ -v^{2}n Q(Q v_g + 2
\Omega)],
$$
$$
 h_{l,n}=-2 n
A_{l}\Omega + B_{l}nQ - 3 n^2 {\Omega}^{2} +v^{2}  n^2 Q^{2},
$$
$$
H_{l,n}=  Q^{2} [-A_{l}   v_g^{2} + B_{l}   v_g
 -3n\Omega  {v_g}^{2}+ v^{2} n ( 2Q v_g +\Omega)],
\end{equation}
$$
A_{l}=2l\omega,\;\;\;\;\;\;\;\;\;\;\;\;\;\;\;\;B_{l}=2lkv^{2}.
$$
The dispersion law of the carrier wave in medium is
\begin{equation}\label{uq19}
\om^{2}=v^{2}k^2 .
\end{equation}

To determine the values of $\varphi_{l,n}^{(\alpha)}$, we equate to zero the
various terms corresponding to the same powers of $\varepsilon$. As a
result, we obtain a chain of equations. Starting with first order in $\ve$, we have
\begin{equation}\label{uq20}
\sum_{l=\pm1}\sum_{n=\pm l} Z_{l} Y_{l,n}\tilde{W}_{l,n}\varphi_{l,n}^{(1)}=0.
\end{equation}
In what follows, we shall also be interested in localized solitary waves which vanish as $ t\rightarrow \pm \infty $. Consequently, according to Eq.\eqref{uq20}, only the following components of $\varphi_{l,n}^{(1)}$ can differ from zero: $\varphi _{\pm 1,\pm 1}^{(1)}$ or $\varphi _{\pm 1,\mp 1}^{(1)}$.
The relations between the parameters $\Omega $ and $Q$ is determined from Eq.\eqref{uq20} and has the form
\begin{equation}\label{e16}
A_{l}n\Omega^{2} -B_{l}n Q \Omega +  \Omega^{3} -v^{2}
Q^{2}\Omega- ln\alpha^{2}_{0} =0.
\end{equation}

To second order in $\ve$ we obtain the relation for $\varphi_{l,n}^ {(2)}$
\begin{equation}\nonumber\\
\sum_{l=\pm1}\sum_{n=-\infty}^{+\infty} Z_{l} Y_{l,n}( \tilde{W}_{l,n}\varphi_{l,n}^ {(2)} +\ve J_{l,n}\frac{\partial \varphi_{l,n}^ {(1)}}{\partial \zeta})=0.
\end{equation}

Substituting  Eq.\eqref{e16} into Eq. \eqref{e15}, we easily see that the following relation holds $J_{\pm 1,\pm 1}=J_{\pm 1,\mp 1}=0$. From the Eq.\eqref{teq17}, to third order in $\ve $, we obtain the following nonlinear equation
\begin{equation}\label{wq17}
i  \frac{\partial \varphi_{+1,\pm 1}^ {(1)}}{\partial \tau}-
 \frac{H_{+1,\pm 1}}{h_{+1,\pm 1}} \frac{\partial^{2} \varphi_{+1,\pm 1}^ {(1)}}{\partial
\zeta^{2}}- \frac{\beta^{2}_{0}}{2 h_{+1,\pm 1}}[
 |\varphi_{+1,\pm 1 }^ {(1)}|^{2}+$$$$(1  - \frac{\Omega_{\mp } }{\Omega_{\pm}})|\varphi_{+1,\mp 1}^ {(1)}|^{2} ] \varphi_{+1,\pm 1}^ {(1)}=0
\end{equation}
where
\begin{equation}\label{67}
H_{l,n}= n Q^{2}\Omega[v^{2} - v^{2}_{g}
 + ln v^{2}_{g} \frac{\alpha^{2}_{0}}{\Omega^{3}} ]
$$$$
h_{l,n}=-2\Omega (ln \omega   +  \Omega
 + ln  \frac{\alpha^{2}_{0}}{2\Omega^{2}})
$$$$
v_g=\frac{B_{l}n  +2v^{2} Q }{A_{l}n + 2  \Omega
+ln \frac{\alpha^{2}_{0}}{\Omega^{2}}}
\end{equation}

From the condition $\hat{E}_{l}=\hat{E}_{-l}^{\ast}$ follows, that
\begin{equation}\label{21}
 {\varphi^{*}}_{l,n}^ {(\alpha)}= \varphi_{-l,-n}^ {(\alpha)}.
\end{equation}

From Eq.\eqref{wq17}, we finally obtain two coupled NLS equations  for functions $U={\ve}\varphi_{+1,+1}^{(1)}$ and $W={\ve}\varphi_{+1,-1}^{(1)}$ that describe the coupling between two components of the pulse
\begin{equation}\label{eq1w}
 i  (\frac{\partial U}{\partial t}+ v_{1} \frac{\partial U}{\partial z})
+  p_{1} \frac{\partial^{2} U
}{\partial z^{2}} + g_{1} |U|^{2}U +r_{1} |W|^{2} U=0,
$$
$$
 i  (\frac{\partial W}{\partial t}+ v_{2} \frac{\partial W}{\partial z})
+  p_{2} \frac{\partial^{2} W
}{\partial z^{2}} + g_{2} |W|^{2}W + r_{2}|U|^{2} W=0,
\end{equation}
where
\begin{equation}\label{eq12w}
v_{1}=v_{{g}_{+1,+1}},\;\;\;\;\;\;\;\;\;\;\;\;\;v_{2}=v_{{g}_{+1,-1}},
$$
$$
p_{1}=\frac{H_{+1,+1}}{-h_{+1,+1}Q^{2}},\;\;\;\;\;\;\;\;\;\;\;\;\;\;\;p_{2}=\frac{H_{+1,-1}}{-h_{+1,-1}Q^{2}},
$$
$$
g_{1}=\frac{ \beta^{2}_{0}}{-2 h_{+1,+1}},\;\;\;\;\;\;\;\;\;\;\;\;\;\;\;\;g_{2}=\frac{ \beta^{2}_{0}}{-2 h_{+1,-1}},
$$
$$
r_{1}=\frac{ \beta^{2}_{0}}{-2 h_{+1,+1}}(1  - \frac{\Omega_{-1} }{\Omega_{+1}}),\;\;\;r_{2}=\frac{ \beta^{2}_{0}}{-2 h_{+1,-1}}(1-\frac{ \Omega_{+1}}{\Omega_{-1}}).
$$
$$
\Omega_{+1}=\Omega_{l=\pm1,n=\pm1},\;\;\;\;\;\;\;\;\;\Omega_{-1}=\Omega_{l=\pm1,n=\mp1}
\end{equation}

The nonlinear equations \eqref{eq1w}  describes the slowly varying envelope functions $U$ and $W$,
where $U$ describes the envelope wave of the frequency $\om+\Om_{+1}$ and $W$ describes the wave with frequency $\om-\Om_{-1}$. The nonlinear coupling between the two waves is governed by the terms $r_{1}|W|^{2}U$ and $r_{2}|U|^{2}W$. We must consider interaction of these field components at different frequencies and solve simultaneously  a set of coupled NLS equations \eqref{eq1w}. A shape-preserving solution of the equations \eqref{eq1w} is a vector pulse because of its two-component structure.

The simplest way to ensure the steady-state property is to require the field envelope functions to depend on the time and space coordinate only through the coordinate $ \xi = t- \frac{z}{V_{0}}$, where $V_{0}$ is the constant pulse velocity.
We are looking  for  solitary-wave solutions for complex amplitudes in the form
\begin{equation}\label{eq12}
U(z,t)=A_{1} S( \xi )e^{i\phi_{1}},\;\;\;\;\;W(z,t)=A_{2}S( \xi )e^{i\phi_{2}},
\end{equation}
where $\phi_{1,2}=k_{1,2} z- \omega_{1,2} t$ are the phase functions, $A_{1},\;$$ A_{2},$$\;k_{1},\;$$k_{2},\;\omega_{1}$ and $\omega_{2}$ are all real constants. Derivatives of the phase $\phi_{1,2}$ are assumed to be small, i.e. the functions $ e^{i\phi_{1,2}}$ are  slow in comparison with oscillations of the pulse and consequently, the inequalities
\begin{equation}\label{eq12a}
k_{1,2}<<Q_{\pm 1},\;\;\;\;\omega_{1,2}<<{\Omega}_{\pm 1}
\end{equation}
are satisfied.

Substituting Eqs.\eqref{eq12}  into Eqs.\eqref{eq1w}  we obtain the nonlinear ordinary differential equation,
\begin{equation}\label{eq14}
(\frac{d S}{d \xi })^{2}=T^{-2}S^{2}-b^{2}S^{4}
\end{equation}
and the relations between quantities $A_{1},A_{2}$ and $\omega_{1},\omega_{2}$ respectively

\begin{equation}\label{rt16}
A_{1}^{2}=\frac{p_{1}q_{2}- p_{2}r_{1}}{p_{2}q_{1}-p_{1}r_{2}}A_{2}^{2},
$$
$$
\omega_{1}=\frac{p_{1}}{p_{2}}\omega_{2}+\frac{V^{2}_{0}(p_{2}^{2}-p_{1}^{2})+v_{2}^{2}p_{1}^{2}-v_{1}^{2}p_{2}^{2}
}{4p_{1}p_{2}^{2}},
\end{equation}
where
\begin{equation}\label{er17}
T^{-2}=V_{0}^{2}\frac{v_{1}k_{1}+k_{1}^{2}p_{1}-\omega_{1}}{p_{1}},\;\;\;
b^{2}=V_{0}^{2} \frac{A_{1}^{2}q_{1}+A_{2}^{2}r_{1}}{2p_{1}},
$$
$$
k_{1}=\frac{V_{0}-v_{1}}{2p_{1}},\;\;\;\;\;\;\;\;\;\;k_{2}=\frac{V_{0}-v_{2}}{2p_{2}}.
\end{equation}

Under the boundary condition  $ S(|\xi| \rightarrow\infty)\rightarrow 0$,  the result of integration of the equation \eqref{eq14} is
 \begin{equation}\label{eq16}
S(\xi)=\frac{1}{bT}sech(\frac{t- \frac{z}{V_{0}}}{T}).
\end{equation}
This is a well known soliton solution.

Hence the two components of the vector soliton $U$ and $W$ are hyperbolic secants $S(\xi)$ with different amplitudes $A_{1,2}$ and the phase functions $\phi_{1,2}$.

Substituting the soliton solution for the function $S(\xi)$ Eq.\eqref{eq16}  of the coupled NLS equations \eqref{eq1w}  into Eqs.\eqref{eq1}  and \eqref{rty32}, we obtain for the x component of the electric field strength $E_{x}(z,t)$ the double periodic vector pulse (vector pulsing soliton) solution of the Maxwell-Bloch equations  \eqref{equ5} and \eqref{eq7}:
\begin{equation}\label{eq17}
E_{x}(z,t)=
\frac{2}{b T}Sech(\frac{t-\frac{z}{V_{0}}}{T})\{ (\Omega_{+1}+\omega_{1})   A_{1} \sin[(k+Q_{+1}+k_{1})z -(\om +\Omega_{+1}+\omega_{1}) t]
$$$$
-(\Omega_{-1}-\omega_{2})A_{2}\sin[(k-Q_{-1}+k_{2})z -(\om -\Omega_{-1}+\omega_{2})t]\}
\end{equation}
The appearance in expression \eqref{eq17}  of the functions  $ \sin[(k+Q_{+1}+k_{1})z -(\om +\Omega_{+1}+\omega_{1}) t]$ and $\sin[(k-Q_{-1}+k_{2})z -(\om -\Omega_{-1}+\omega_{2})t]$ indicates the formation of double periodic beats with coordinate and time relative to the frequency and wave number of the carrier wave ($\omega$, $k$), with  characteristic parameters ($\om +\Omega_{+1}$, $k+Q_{+1}$) and ($\om -\Omega_{-1}$, $k-Q_{-1}$), respectively ( taking into account the Eq. \eqref{eq12a}), as a result of which the soliton solutions \eqref{eq16}  for $S(\xi)$ is transformed into vector pulsing soliton solution \eqref{eq17}  for the $x$ component of the strength of the electric field of the optical pulse $E_{x}(z,t)$. Eq.\eqref{eq17} is exact regular time and space double periodic solution of nonlinear equation \eqref{equ5} and \eqref{eq7} which, like soliton and breather, loses no energy in the process of propagation through medium on the considerable distances.

\vskip+0.5cm

\centerline{IV. Conclusion}

We have shown that in the propagation of optical pulse through resonance medium containing an ensemble of SQDs under the condition of SIT  optical vector pulsing soliton can arise. The explicit form and parameters of the optical vector pulsing soliton are given by Eqs.\eqref{eq17}, \eqref{67}, \eqref{eq12w}, \eqref{rt16} and \eqref{er17}. The dispersion equation and the relations between quantities $\Omega_{\pm 1}$ and $Q_{\pm 1}$ are given by Eqs.\eqref{uq19} and \eqref{e16}, respectively.

The conditions of the existence Eqs. \eqref{rtyp} and \eqref {eq12a}  of a vector pulsing soliton Eq.\eqref{eq17} with phase modulation  is connected. Indeed, from the Eqs. \eqref{e16} and \eqref{21} it is  clear that for the pulse free of phase modulation the expansion Eq. \eqref{rty32} is not valid  and for such pulse  other approaches are used (see Ref.\cite {Adamshvili:Result:11} and references therein). In the last case for the existence of the vector pulse, besides   Eqs.\eqref{rtyp} and \eqref {eq12a}, will be necessary to satisfy some other additional conditions. In addition,  a vector pulse  exists in the region of frequencies which on two- or three order less in comparison with carrier frequency \cite {Adamshvili:Result:11}.

Using typical parameters for the pulse, the materials, and the SQDs \footnote{Parameters for the numerical simulation: $\omega =6\times 10^{15} {\rm Hz}$, $T=2$ps,  $\mu_{0} = 2\times10^{-17} \, {\rm esu \, cm}$, $n_0 = \times 10^{10} \, {\rm cm^{-3}}$,  $\eta = 3.3$, full-width half-maximum inhomogeneous broadening $\hbar \delta^* = 60 \, {\rm meV}.$}
we can construct a plot of the $x$ component of the electric field for the double periodic vector pulsing soliton Eq.\eqref{eq17} (shown in Fig.1 for a fixed value of the $z$ coordinate). The coupled NLS equations \eqref{eq1w} represent two-component vector pulsing soliton whose shape we can obtain from the Eq.\eqref{eq17}. Both components of a vector pulsing soliton are bright breathers, because the conditions $p_{1}g_{1}>0$ and $p_{2}g_{2}>0$, are fulfilled. Consequently, for the parameters [17], this case corresponds to a two-component bright vector pulsing soliton.

The breather is the special case of the vector pulsing soliton when one of the amplitudes $A_{1}$ or $A_{2}$ is equal to zero and  Eq.\eqref{eq17} is valid only for one of the pair of parameters $(\om+\Omega_{+1},\;k+Q_{+1})$  or $(\om-\Omega_{-1},\;k-Q_{-1})$.  The profile of the breather with the same value of the parameters as for a vector pulsing soliton (at $A_{2}=0$), for a fixed value of $z=0$, on the Fig.2 is presented. The shape of a vector pulsing soliton (Fig.1) is different in the comparison with the shape of the scalar breather (Fig.2).

Our results show that a distribution of the transition dipole moments in an ensembles of SQDs significantly changes  the polarization Eq.\eqref{eqy8} and Maxwell's wave equation \eqref{rty2}.   On taking into account a distribution of the dipole moments in an ensembles of SQDs, we conclude that the  number of effectively active quantum dots is reduced by the dipole distribution if the total number of SQDs is kept constant and this leads to a decrease of the absorption of the wave energy during propagation in the medium. From the Eqs. \eqref{e16}, \eqref{67},\eqref{eq12w}, \eqref{rt16} and \eqref{er17},  it is clear that the parameters of a vector pulsing soliton contains the quantities $\alpha^{2}_{0}$ and $\beta^{2}_{0}$ [Eq. \eqref{teu17}] and consequently, depend on the fluctuations of the dipole moments of SQDs.

In conclusion, we have predicted the existence of a vector breathers in a sample of inhomogeneously broadened and of the fluctuations of the dipole moments, a three-level SQDs in the presence of single-excitonic and biexcitonic transitions  for experimentally relevant parameters. We hope that the presented results will stimulate the research of multicomponent vector pulses in quantum dot nanostructures due to their importance for the lossless and shape-invariant transport of information on nanoscales. It should be noted that the constructed theory is quite general and can be simplified for atomic systems, for example, for dielectrics with optically active two-level impurities, whose the dipole moments are constants.

\begin{figure}
\begin{centering}
\includegraphics{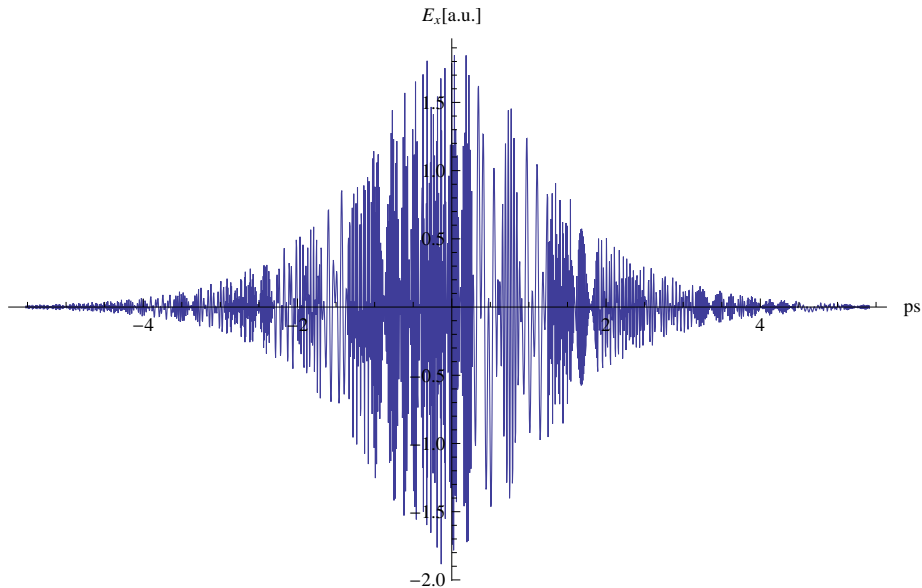}
\end{centering}
\caption{ The strength of the optical electrical field $E_{x}(0,t)$ the vector pulsing soliton is shown for a fixed value of $z$. The nonlinear pulse  oscillates with the sum  $\om+\Om_{+1}+\om_{1}$ and difference  $\om-\Om_{-1}+\om_{2}$   of the frequencies  along the $t$-axis. }
\label{fig1}
\end{figure}
\begin{figure}
\begin{centering}
\includegraphics{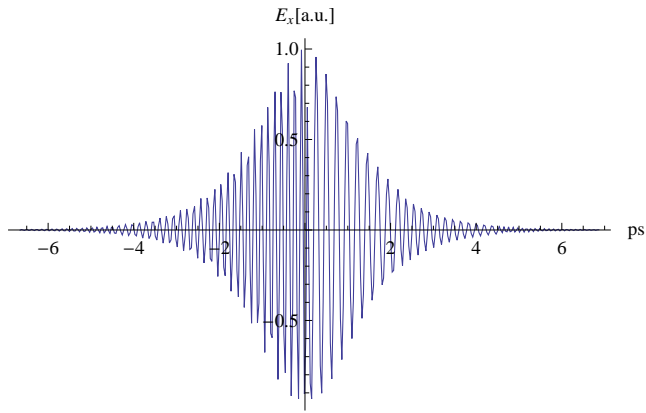}
\end{centering}
\caption{The strength of the optical electrical field $E_{x}(0,t)$ the breather is shown for a fixed value of $z$.
The breather  oscillates with the frequency  $\om+\Om_{+1}+\om_{1}$ along the $t$-axis. }
\label{fig2}
\end{figure}


\begin{thebibliography}{20}
\expandafter\ifx\csname natexlab\endcsname\relax\def\natexlab#1{#1}\fi
\expandafter\ifx\csname bibnamefont\endcsname\relax
  \def\bibnamefont#1{#1}\fi
\expandafter\ifx\csname bibfnamefont\endcsname\relax
  \def\bibfnamefont#1{#1}\fi
\expandafter\ifx\csname citenamefont\endcsname\relax
  \def\citenamefont#1{#1}\fi
\expandafter\ifx\csname url\endcsname\relax
  \def\url#1{\texttt{#1}}\fi
\expandafter\ifx\csname urlprefix\endcsname\relax\def\urlprefix{URL }\fi
\providecommand{\bibinfo}[2]{#2}
\providecommand{\eprint}[2][]{\url{#2}}


\bibitem[{\citenamefont{Bimberg}(1999)}]{Bimberg::1999}
\bibinfo{author}{\bibfnamefont{D.}~\bibnamefont{Bimberg}},
\bibinfo{author}{\bibfnamefont{M.}~\bibnamefont{Grundmann}}\bibnamefont{and}
\bibinfo{author}{\bibfnamefont{N.~N.}~\bibnamefont{Ledentsov}},
 \bibinfo{journal}{Quantum Dot Heterostructures,(Wiley, Chichester,)} (\bibinfo{year}{1999}).



\bibitem[{\citenamefont{Borri et~al.}(2001)\citenamefont{Borri, Langbein,
  Schneider, Woggon, Sellin, Ouyang, and Bimberg}}]{Borri::02}
\bibinfo{author}{\bibfnamefont{P.}~\bibnamefont{Borri}},
  \bibinfo{author}{\bibfnamefont{W.}~\bibnamefont{Langbein}},
  \bibinfo{author}{\bibfnamefont{S.}~\bibnamefont{Schneider}},
  \bibinfo{author}{\bibfnamefont{U.}~\bibnamefont{Woggon}},
  \bibinfo{author}{\bibfnamefont{R.~L.} \bibnamefont{Sellin}},
  \bibinfo{author}{\bibfnamefont{D.}~\bibnamefont{Ouyang}}, \bibnamefont{and}
  \bibinfo{author}{\bibfnamefont{D.}~\bibnamefont{Bimberg}},
  \bibinfo{journal}{Phys. Rev. B} \textbf{\bibinfo{volume}{66}},
  \bibinfo{pages}{081306} (\bibinfo{year}{2002}).

\bibitem[{\citenamefont{Panzarini et~al.}(2002)\citenamefont{Panzarini,
 Hohenester, and Molinari}}]{Panzarini:PhysRevB:02}
\bibinfo{author}{\bibfnamefont{G.}~\bibnamefont{Panzarini}},
  \bibinfo{author}{\bibfnamefont{U.}~\bibnamefont{Hohenester}},
  \bibnamefont{and} \bibinfo{author}{\bibfnamefont{E.}~\bibnamefont{Molinari}},
  \bibinfo{journal}{Phys. Rev. B} \textbf{\bibinfo{volume}{65}},
  \bibinfo{pages}{165322} (\bibinfo{year}{2002}).


 \bibitem[{\citenamefont{Adamshvili et~al.}(2008)\citenamefont{Adamshvili,
  }}]{AdamshviliMaradudin:Arxiv:2010}
\bibinfo{author}{\bibfnamefont{G.~T.} \bibnamefont{Adamshvili}} \bibnamefont{and}
  \bibinfo{author}{\bibfnamefont{A.~A.}~\bibnamefont{Maradudin,}}
    \bibinfo{journal}{Preprint, Arxiv,}  \textbf{1003.0638v1.},
 (\bibinfo{year}{2 Mar 2010}).


\bibitem[{\citenamefont{Allen and Eberly}(1975)}]{Allen::75}
\bibinfo{author}{\bibfnamefont{L.}~\bibnamefont{Allen}} \bibnamefont{and}
  \bibinfo{author}{\bibfnamefont{J.}~\bibnamefont{Eberly}},
  \emph{\bibinfo{title}{Optical resonance and two level atoms}}
  (\bibinfo{publisher}{Dover}, \bibinfo{year}{1975}).


\bibitem[{\citenamefont{Chen et~al.}(2008)\citenamefont{Chen}}]{Chen:Phys. Rev. E :04}
\bibinfo{author}{\bibfnamefont{M.} \bibnamefont{Chen}},
  \bibinfo{author}{\bibfnamefont{D.~J.}~\bibnamefont{Kaup}} \bibnamefont{and}
  \bibinfo{author}{\bibfnamefont{B.~A.}~\bibnamefont{Malomed,}}
    \bibinfo{journal}{Phys. Rev. E }  \textbf{\bibinfo{volume}{69}},
 \bibinfo{pages}{056605 } (\bibinfo{year}{2004}).



\bibitem[{\citenamefont{Giessen et~al.}(1998)\citenamefont{Giessen, Knorr,
  Haas, Koch, Linden, Kuhl, Hetterich, Gr\"un, and
  Klingshirn}}]{Giessen:PhysRevLett:98}
\bibinfo{author}{\bibfnamefont{H.}~\bibnamefont{Giessen}},
  \bibinfo{author}{\bibfnamefont{A.}~\bibnamefont{Knorr}},
  \bibinfo{author}{\bibfnamefont{S.}~\bibnamefont{Haas}},
  \bibinfo{author}{\bibfnamefont{S.~W.} \bibnamefont{Koch}},
  \bibinfo{author}{\bibfnamefont{S.}~\bibnamefont{Linden}},
  \bibinfo{author}{\bibfnamefont{J.}~\bibnamefont{Kuhl}},
  \bibinfo{author}{\bibfnamefont{M.}~\bibnamefont{Hetterich}},
  \bibinfo{author}{\bibfnamefont{M.}~\bibnamefont{Gr\"un}}, \bibnamefont{and}
  \bibinfo{author}{\bibfnamefont{C.}~\bibnamefont{Klingshirn}},
  \bibinfo{journal}{Phys. Rev. Lett.} \textbf{\bibinfo{volume}{81}},
  \bibinfo{pages}{4260} (\bibinfo{year}{1998}).

\bibitem[{\citenamefont{Schneider et~al.}(2003)\citenamefont{Schneider, Borri,
  Langbein, Woggon, F\"orstner, Knorr, Sellin, Ouyang, and
  Bimberg}}]{Schneider:ApplPhysLett:03}
\bibinfo{author}{\bibfnamefont{S.}~\bibnamefont{Schneider}},
  \bibinfo{author}{\bibfnamefont{P.}~\bibnamefont{Borri}},
  \bibinfo{author}{\bibfnamefont{W.}~\bibnamefont{Langbein}},
  \bibinfo{author}{\bibfnamefont{U.}~\bibnamefont{Woggon}},
  \bibinfo{author}{\bibfnamefont{J.}~\bibnamefont{F\"orstner}},
  \bibinfo{author}{\bibfnamefont{A.}~\bibnamefont{Knorr}},
  \bibinfo{author}{\bibfnamefont{R.~L.} \bibnamefont{Sellin}},
  \bibinfo{author}{\bibfnamefont{D.}~\bibnamefont{Ouyang}}, \bibnamefont{and}
  \bibinfo{author}{\bibfnamefont{D.}~\bibnamefont{Bimberg}},
  \bibinfo{journal}{Appl. Phys. Lett.} \textbf{\bibinfo{volume}{83}},
  \bibinfo{pages}{3668} (\bibinfo{year}{2003}).

 \bibitem[{\citenamefont{Adamashvili and Knorr}(2006)}]{Adamashvili:OptLett:06}
\bibinfo{author}{\bibfnamefont{G.}~\bibnamefont{Adamashvili}} \bibnamefont{and}
  \bibinfo{author}{\bibfnamefont{A.}~\bibnamefont{Knorr}},
  \bibinfo{journal}{Opt. Lett.} \textbf{\bibinfo{volume}{31}},
  \bibinfo{pages}{74} (\bibinfo{year}{2006}).

\bibitem[{\citenamefont{Adamshvili et~al.}(2007)\citenamefont{Adamshvili,
  Weber, Knorr, and Adamshvili}}]{Adamshvili:PhysRevA:07}
\bibinfo{author}{\bibfnamefont{G.~T.} \bibnamefont{Adamshvili}},
  \bibinfo{author}{\bibfnamefont{C.}~\bibnamefont{Weber}},
  \bibinfo{author}{\bibfnamefont{A.}~\bibnamefont{Knorr}}, \bibnamefont{and}
  \bibinfo{author}{\bibfnamefont{N.~T.} \bibnamefont{Adamshvili}},
  \bibinfo{journal}{Phys. Rev. A},  \textbf{\bibinfo{volume}{75}},
  \bibinfo{pages}{063808} (\bibinfo{year}{2007}).

\bibitem[{\citenamefont{Adamashvili and  Knorr}(2007)}]{Adamashvili:PhysLettA:07}
\bibinfo{author}{\bibfnamefont{G.~T.} \bibnamefont{Adamashvili}}
  \bibnamefont{and} \bibinfo{author}{\bibfnamefont{A.}~\bibnamefont{Knorr}},
  \bibinfo{journal}{Phys. Lett. A},  \textbf{\bibinfo{volume}{367}},
  \bibinfo{pages}{220} (\bibinfo{year}{2007}).

 \bibitem[{\citenamefont{Adamshvili et~al.}(2008)\citenamefont{Adamshvili,
  Weber, Knorr, and Adamshvili}}]{Adamshvili:Eur.Phys.J.D.:08}
\bibinfo{author}{\bibfnamefont{G.~T.} \bibnamefont{Adamshvili}},
  \bibinfo{author}{\bibfnamefont{C.}~\bibnamefont{Weber}} \bibnamefont{and}
  \bibinfo{author}{\bibfnamefont{A.}~\bibnamefont{Knorr,}}
    \bibinfo{journal}{The Eur. Phys. J. D}  \textbf{\bibinfo{volume}{47}},
 \bibinfo{pages}{113} (\bibinfo{year}{2008}).

\bibitem[{\citenamefont{Adamshvili}(2009)\citenamefont{Adamshvili }}]{Adamashvili:Opt.and Sprct:09}
\bibinfo{author}{\bibfnamefont{G.~T.} \bibnamefont{Adamshvili}},
\bibinfo{author}{\bibfnamefont{N.~T.} \bibnamefont{Adamshvili}},
\bibinfo{author}{\bibfnamefont{M.~D.} \bibnamefont{Peikrishvili}},
\bibinfo{author}{\bibfnamefont{G.~N.} \bibnamefont{Motsonelidze}} \bibnamefont{and}
\bibinfo{author}{\bibfnamefont{R.~R.} \bibnamefont{Koplatadze}},
   \bibinfo{journal}{Optics and Spectroskopy}  \textbf{\bibinfo{volume}{106}},
  \bibinfo{pages}{972} (\bibinfo{year}{2009}).


\bibitem[{\citenamefont{Adamshvili et~al.}(2007)\citenamefont{Adamshvili}}]{Adamshvili:Result:11}
\bibinfo{author}{\bibfnamefont{G.~T.} \bibnamefont{Adamshvili}},
    \bibinfo{journal}{Results in  Physics},  \textbf{\bibinfo{volume}{1}},
  \bibinfo{pages}{26} (\bibinfo{year}{2011}).

\bibitem[{\citenamefont{Landau }(1980)}]{Landau:Quantum Mechanics:80}
\bibinfo{author}{\bibfnamefont{L.~D.}~\bibnamefont{Landau }} \bibnamefont{and}
\bibinfo{author}{\bibfnamefont{E.~M.}~\bibnamefont{Lifshitz}}
\emph{\bibinfo{title}{Quantum Mechanics, Nonrelativictic theory.}}
(\bibinfo{publisher}{Pergamon press Ltd. p.544}, \bibinfo{year}{1980}).


\bibitem[{\citenamefont{ Taniuti}(1973)}]{Taniuti::1973}
\bibinfo{author}{\bibfnamefont{T.}\bibnamefont{Taniuti}} \bibnamefont{and}
  \bibinfo{author}{\bibfnamefont{N. }\bibnamefont{Iajima}},
  \bibinfo{journal}{J. Math. Phys.} \textbf{\bibinfo{volume}{14}},
  \bibinfo{pages}{1389} (\bibinfo{year}{1973}).







\end{thebibliography}
\end{document}